# Locally Frozen Magnetic Field in HTSC Ceramics

S. I. Bondarenko, A. A. Shablo, V. P. Koverya, and D. Yu. Fomin

*Verkin Institute for Low Temperature Physics and Engineering, National Academy of Sciences of Ukraine, Kharkov, 61103 Ukraine*

*e-mail: bondarenko@ilt.kharkov.ua*

**Abstract**—The value of a locally frozen magnetic field in a region with a diameter of 0.5 mm in a 0.5-mm-thick $YBa_2Cu_3O_{7-x}$ plate was investigated as a function of the excitation field (to $2 \times 10^4$ A m$^{-1}$), plate cooling mode (in the absence or presence of a field; i.e., zero-field cooling (ZFC) or field coupling (FC)), and local demagnetizing field. Analysis of the measurement results in the noted range of excitation fields showed the following: (i) the dependence on the excitation field for the ZFC mode is explained by the local inhomogeneity of critical currents of weak links in the ceramic Josephson medium and is limited by their maximum value at the temperature of the experiment (77 K); (ii) the dependence on the excitation field for the FC mode contains a portion of the magnetic phase transition from the frozen current structure, typical of the initial portion of the dependence, to the current structure characteristic of the ZFC freezing mode, and is limited by this transition; and (iii) the dependence on the demagnetizing field for the ZFC mode can be explained by the stable coexistence (without annihilation) of microscopic current loops with opposite current directions in the ceramics.

## INTRODUCTION

In recent years, several research teams have begun investigation of the methods for forming local (both microscopic and macroscopic) regions of frozen magnetic field in high-temperature superconductors (HTSCs) and the properties of these regions [1–4]. The formation of such a region is equivalent to local magnetization of a superconducting sample, because the latter acquires an additional magnetic moment after the field freezing. Such a frozen magnetic field and the corresponding frozen one- or multiquantum magnetic flux should be supported by a vortex current structure with a particular degree of complexity.

The interest in this line of research is caused, in particular, by the possibility of direct measurement of the pinning force of a single Abrikosov vortex [1] and a multiphoton single macroscopic vortex [4], as well as the possibility of physical simulation of the static and dynamic properties of various current structures [2, 3], both single ones and lattices formed in a controlled way.

Field freezing in superconductors was generally implemented in a uniform (i.e., nonlocal) field. The thus frozen magnetic field in HTSC materials is characterized by two specific features: (i) it may change (relax) with time and (ii) the critical penetration field into a sample depends on the height of the edge barrier. The latter, in turn, is determined by the technique of barrier formation. Both these features complicate methods for study and decrease their accuracy. Our investigations [5] have shown that a locally frozen field in HTSC ceramics does not change in time (i.e., does not relax), and the edge barrier is absent due to the geometry of the frozen magnetic field. These characteristics of the local field freezing are favorable for wider application of the noted method for studying superconductors.

Field freezing is performed using one of the two known modes of sample cooling to a temperature below the superconductor's critical temperature ($T_C$). In the zero-field cooling (ZFC) mode, a sample is cooled below $T_C$ in a zero field, after which an external uniform excitation field (exceeding the critical value) is applied to the sample with subsequent switching off. In the field cooling (FC) mode, a sample is cooled in an external field, and the excitation field is switched off after the sample cooling is over. There is no complete understanding of the following questions: (i) how the local excitation field in the above-mentioned modes affects the local frozen magnetic field in a Josephson ceramic medium when the ceramic grains are not involved in freezing? and (ii) how stable the local frozen magnetic field is to the action of an additional external local dc magnetic field? The stability data may be important, in particular, in consideration of a superconductor as a carrier of magnetic information [6] that must be not only recorded but also erased.

In this paper, we report the results of investigation of the noted questions by the example of $YBa_2Cu_3O_{7-x}$ ceramics with weak Josephson links between grains, which are typical of ceramics with a low critical current density (30 A cm$^{-2}$ at 77 K for the samples under study), obtained from the current–voltage characteristic of the sample, at local excitation fields for the frozen magnetic field that are much stronger (up to $2 \times 10^4$ A m$^{-1}$) than those used by us previously [4] but do not exceed the first critical field of ceramic grains at 77 K.

## STATEMENT OF EXPERIMENTS

The experiments were performed in a cryostat with liquid nitrogen. The measuring cell (Fig. 1) is immersed in liquid nitrogen and contains a generator of local excitation field $H_e$, detector of frozen magnetic field, and a ceramic sample in the form of a square plate with a side of 9 mm and thickness of 0.5 mm. The critical temperature and width of the superconducting transition in the ceramics studied are, respectively, $T_C = 89$ K and $\Delta T = 2$ K. The sample is connected by a rod with a micrometer screw located on the cryostat flange and can be displaced in the vertical direction in the cell. The source of the field $H_e$ is a microsolenoid with a diameter of 0.5 mm and a small (0.7 mm) gap in the central part to provide sample displacement. The frozen magnetic field detector is a flux gate magnetometer [7] with a uniform-field sensitivity equal to $10^{-3}$ A m$^{-1}$. The diameter of the detector end part, adjacent to the sample surface, is 1.5 mm and the detector length is ~10 mm. Rotation of the cryostat around its vertical axis makes it possible to orient the sample plane parallel to the horizontal component of the Earth's magnetic field. In this case, the component of this field that is perpendicular to the sample plane is zero (accurate to 0.1 A m$^{-1}$), a fact that makes it possible to exclude the Earth's field on the currents excited by the local excitation field, which is perpendicular to the sample surface. Using a microsolenoid, one can generate for a long time (without the microsolenoid overheating) a local excitation field up to $10^4$ A m$^{-1}$ (for the FC mode) at any point of the sample along the $X$ axis, and, for a short time, a field stronger by a factor of 2 (for the ZFC mode).

To study the local frozen magnetic field formed in the ZFC mode, the sample was cooled to 77 K in the absence of excitation field and with the effect of the Earth's magnetic field excluded in the way described above. Then a specified constant excitation field was formed in the center of the sample during 5–10 s using a microsolenoid. After switching off the field, the sample center could be displaced in the detector region and the maximum value $H_{fm}$ of the vertical component of the frozen magnetic field was detected. The distribution $H_f(x)$ of this frozen field component along the sample axis $X$ was measured in the same way. To obtain the dependence $H_f(x)$ at another excitation field, the cell with the sample was lifted above the level of liquid nitrogen into the vapor region of the cryostat, heated at a temperature exceeding $T_C$, and immersed again in liquid nitrogen.

To study the frozen magnetic field in the FC mode, the cell with a sample was installed in the vapor region of the cryostat at $T > T_C$, and the excitation field of the microsolenoid was applied. Then the cell with the sample was immersed in liquid nitrogen and cooled to $T = 77$ K, after which the field was switched off. The dependence $H_f(x)$ and the $H_{fm}$ value were determined in the way described above.

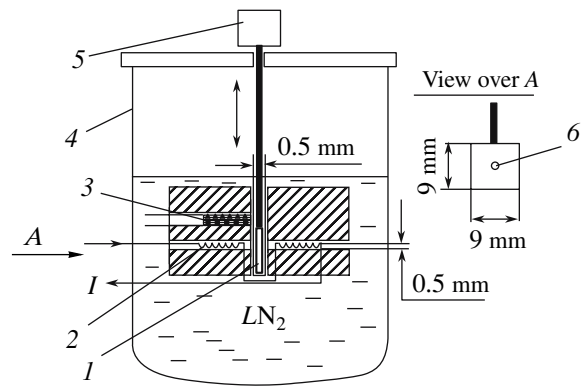

**Fig. 1.** Schematic of the experimental cell located in a liquid-nitrogen cryostat: (*1*) HTSC ceramic plate, (*2*) microsolenoid, (*3*) fluxgate magnetometer sensor, (*4*) Dewar flask, (*5*) micrometer screw, and (*6*) current vortex maintaining the frozen magnetic field in a square ceramics plate with a side of 9 mm.

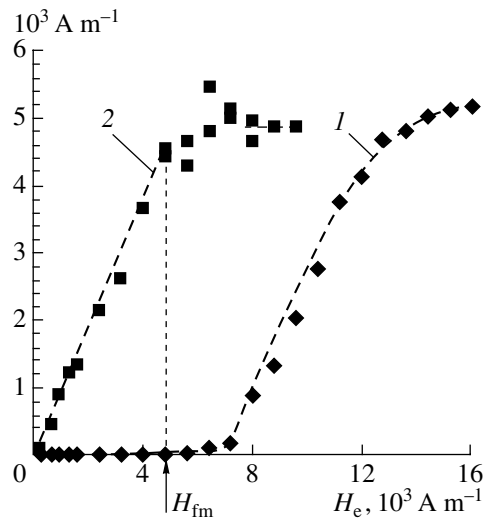

**Fig. 2.** Dependence of the maximum frozen magnetic field ($H_{fm}$) on the excitation field $H_e$ for the (*1*) ZFC and (*2*) FC modes of field freezing.

In the experiments on the study of the stability of the local frozen magnetic field with respect to the additional external local magnetic field $H_a$, the region with a frozen magnetic field in the superconducting ceramics was located in the microsolenoid gap, and, using the microsolenoid, additional constant fields of different orientation and magnitude (below the field $H_e$ corresponding to the formation of the frozen magnetic field) was generated for a short time (5–10 s). Then the changes in $H_f(x)$ and $H_{fm}$ were investigated.

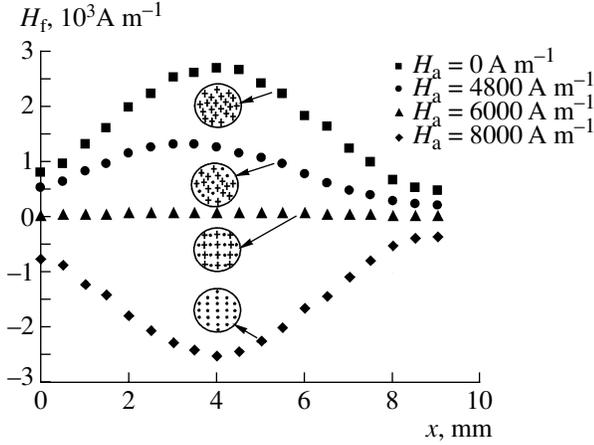

**Fig. 3.** Distribution of the vertical component of the frozen magnetic field along the $x$ axis of the sample after the demagnetizing action of different additional dc fields. The initial field was frozen in the ZFC mode by imposing an excitation field of 8000 A m$^{-1}$. The frozen fields of different directions (+ or ●) in microcircuits in the ceramics are schematically shown (by circles) for the noted dependences.

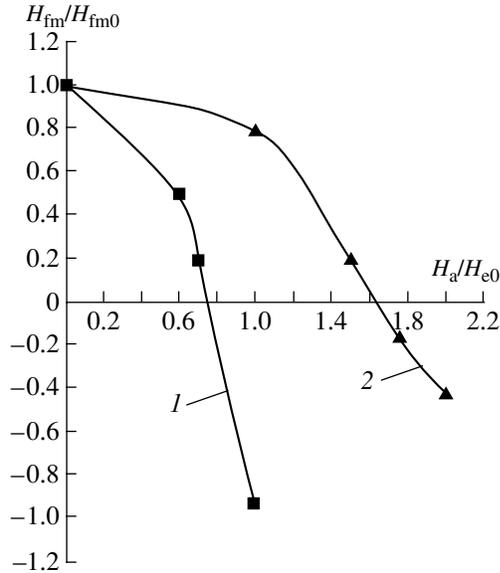

**Fig. 4.** Dependences of the relative value of the maximum frozen magnetic field ($H_{fm0}$ is the largest value of the initial frozen field) on the relative additional demagnetizing field ($H_{e0}$ is the dc excitation field for the initial frozen field): effect of the dc demagnetizing field in the (*1*) ZFC and (*2*) FC modes.

## EXPERIMENTAL RESULTS AND DISCUSSION

Three types of experiments were performed. In the first-type experiments, we determined the dependence $H_{fm}(H_e)$ for the case of field freezing in the ZFC mode (Fig. 2, curve *1*). This dependence contains three regions. In the first (initial) region, $H_{fm} = 0$. The beginning of the second region corresponds to the first signs of frozen magnetic field at some threshold value of $H_e$, which can be referred to as the local critical field $H_{ec}$ corresponding to the field penetration into sample and freezing after switching off the excitation field. In the case under consideration, $H_{ec} \approx 4.8 \times 10^3$ A m$^{-1}$. At $H_e > 4.8 \times 10^3$ A m$^{-1}$, the frozen magnetic field increases proportionally to $H_e$ up to the saturation region of the frozen field ($H_e > 1.3 \times 10^4$ A m$^{-1}$). In the second-type experiments, the dependence $H_{fm}(H_e)$ was determined for the FC mode (Fig. 2, curve *2*). Curve *2* demonstrates three characteristic regions. The first one is almost linear ($H_e = 0$–$4.8 \times 10^3$ A m$^{-1}$), where the frozen field magnitude is similar to that of the excitation field. In the second region ($H_e$ ranges from $4.8 \times 10^3$ to $8 \times 10^3$ A m$^{-1}$), unreplicated values of the frozen field were observed at the same field $H_e$ (unreplicated region). The beginning of the unreplicated region coincides with $H_{ec}$. Finally, the third region (with $H_e$ ranging from $8 \times 10^3$ to $9.6 \times 10^3$ A m$^{-1}$) exhibits "saturation" of the frozen magnetic field. It can also be seen that the values of the frozen field in the saturation regions in curves *1* and *2* are approximately the same. In the experiments of the third type, we analyzed the effect of the additional local constant field $H_a$ of different orientations on the frozen magnetic field formed previously in the ZFC and FC modes. It was established that the additional field oriented in the same direction as the frozen magnetic field does not change the latter. Vice versa, after the application of an additional field oriented in the opposite direction, the frozen field may not only decrease but also change sign (Fig. 3). In this case, the additional field stronger affects the frozen magnetic field formed in the ZFC mode in comparison with the FC mode. As can be seen in Fig. 3, to obtain complete demagnetization ("erasure") of the former field, it is sufficient to have $H_a \approx 0.75 H_e$, whereas the complete reversal of the initial frozen magnetic field occurs at $H_a = H_e$. As a result, a frozen magnetic field is formed that has the same magnitude as the initial one but opposite sign. Erasure of the frozen magnetic field formed in the FC mode requires a field stronger by a factor of 2. This difference is especially pronounced in the dependences of the relative value of the maximum frozen magnetic field (derived from the dependence $H_f(x)$) on the relative value of the additional field (Fig. 4, curves *1*, *2*).

The results of the experiments performed can be explained on the basis of the representation of the HTSC ceramics as a Josephson medium [8]. This medium has the form of a multilayer grid of superconducting microcircuits composed of fine-grained (mainly 0.8–1.2 μm in size) completely superconducting ceramic grains connected by weak Josephson links; these microcircuits have different critical currents. A local magnetic excitation field can generate closed currents of the same direction in microcircuits. Thus, a current microcircuit corresponds to each physical microcircuit.

Let us first consider the local frozen magnetic field formed in the ZFC mode.

In the first-type experiments, the region of the dependence $H_{fm}(H_e)$ at $H_e < H_{ec}$ corresponds to complete shielding of the excitation field by the ceramics. In this case, the relation $\Phi_e = \Phi_I$ is satisfied, where $\Phi_e$ is the magnetic flux of the excitation field and $\Phi_I$ is the flux of the total screening current in the ceramic region located in this field, which has the opposite direction. At a larger $H_e$ value ($H_e = H_{ec}$), the screening current of individual microcircuits with the weakest links reaches the critical value. Such current microcircuits cease screening the field $H_e$. This situation corresponds to the relation $\Phi_e > \Phi_I$. After switching off such a field, according to the flux conservation law for multiply connected superconductors, the magnetic flux $\Phi = \Phi_e - \Phi_I$ is frozen in this region of the ceramics and a weak frozen magnetic field is formed. Thus, a current structure in the form of a grid of current microcircuits, different in shape and size and having small critical currents, corresponds to this type of frozen magnetic field (Fig. 5a). The frozen magnetic field will increase with an increase in the field $H_e$ until the Josephson currents in all microcircuits reach the maximum critical values and cease to screen $H_e$. Switching off the excitation field after this event leads to freezing of the maximum possible (for this region of the ceramics) magnetic field, i.e., to the transition to the saturation region in the dependence $H_{fm}(H_e)$ (Fig. 2, curve $1$).

Thus, the difference between the value $H_e$, corresponding to the onset of saturation of the frozen magnetic field, and $H_{ec}$ can be explained by the inhomogeneity of the ceramic region with frozen field with respect to the critical currents of weak links. Hence, the dependence $H_{fm}(H_e)$ for the ZFC mode can be used for local magnetic diagnostics of the Josephson inhomogeneity of the ceramics. At the same time, the value of the saturated frozen field ($H_{fms}$) can be used to estimate the local maximum density ($j_{cm}$) of the critical current ($I_{cm}$) of such a ceramics. If we use the relations

$$\mu_0 H_{fms} S_f \approx L_I I_{cm}, \qquad (1)$$

$$S_f = \pi r^2, \qquad (2)$$

$$I_{cm} \approx j_{cm} r t, \qquad (3)$$

where $\mu_0 = 4 \times 10^{-7}$ H m$^{-1}$, $S_f$ is the area of the region with frozen magnetic field, $r$ is the radius of the region with frozen field, $t$ is the plate thickness, and $L_I$ is the effective inductance of a current loop with frozen field, we have

$$j_{cm} \approx (\pi \mu_0 r / L_I t) H_{fms}. \qquad (4)$$

Estimation of $j_{cm}$ from formula (4) gives a value that significantly exceeds the value obtained from the current–voltage characteristic of the plate; this discrepancy can be explained by the actual spatial inhomogeneity of the plate and nonuniform current distribution in the plate during measurement of its current–voltage characteristic.

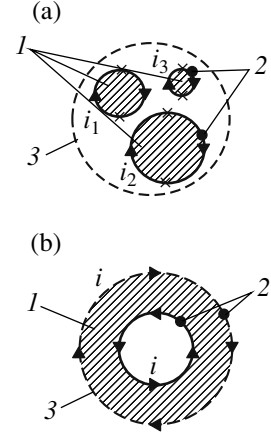

**Fig. 5.** Schematic representation of the current structure of the vortex type supporting the (a) ZFC and (b) FC frozen fields: ($1$) frozen-field regions in the ceramics, ($2$) current loops with different critical currents ($i_1$, $i_2$, $i_3$) of Josephson links (●) for the ZFC mode and the current $i$ for the FC mode, and ($3$) the circle corresponding to the diameter of the microsolenoid generating the external excitation field.

Let us consider a possible mechanism of action of the additional field $H_a$ on the frozen field formed in the ZFC mode and the local current structure corresponding to the frozen field. The field $H_a$ may have the same direction as the frozen magnetic field or opposite. In the first case, the additional field cannot change the frozen field if the relation $H_a \leq H_e$ is satisfied. Upon superposition of the $H_a$ with opposite direction, in some microcircuits, for which the relation $\Phi_e = \mu_0 H_a S_f > L_m i_{mc}$ holds true ($\Phi_e$ is the magnetic flux of the field $H_a$ and $S_f$, $L_m$ and $i_{mc}$ are, respectively, the area, inductance, and critical current of a microcircuit), a flux $\Phi_{fa} = L_m i_{mc}$, with the direction opposite to the frozen field flux, can be frozen. As a result, the total frozen flux ($\mu_0 H_f S_f - \Phi_{fa}$) becomes lower than the flux $\mu_0 H_f S_f$ of the initial frozen magnetic field. The fluxgate will detect a smaller value of the frozen magnetic field. With a further increase in $H_a$, a situation may arise where frozen fluxes of different directions have the same magnitude. In this case, the $H_a$ value can be smaller than the field $H_e$ that formed the initial frozen magnetic field. In this situation, our relatively large magnetic detector with a low spatial resolution will show the absence of the integral frozen magnetic field, although opposite, mutually compensated magnetic fluxes will be frozen in the network of these microcircuits. The current structure of the transformed frozen field is a network of current microcircuits with opposite currents. The complete magnetization reversal of a local region in ceramics with the use of an additional field will occur when this field if equal in magnitude to the excitation field for the initial frozen magnetic field. In this case, the current structure is a network of current microcircuits with currents oriented in the same direction that is opposite to the initial one. As a result, a frozen magnetic field will be formed, equal to the initial field in magnitude but having the opposite direction. This concept was confirmed in the third type of experiments (Fig. 3; Fig. 4, curve *1*). The suggested transformation of the distribution of frozen fluxes of different direction (+ and ●) with increasing $H_a$ is schematically shown in insets in Fig. 3 near the corresponding dependences $H_f(x)$. These dependences were stable (checked for an hour). Hence, one can suggest stable coexistence (without annihilation) of micrometer-scale vortex structures with opposite current direction in the ceramics, in contrast to the annihilation of vortices and antivortices in $YBa_2Cu_3O_{7-x}$ films [3].

Let us now consider the frozen magnetic field formed in the FC mode. The results of the second type of the experiments can be explained if we assume that this field is maintained by a radically different local current structure. Its schematic diagram, justified by us previously in [4], is shown in Fig. 5b. At $H_e < H_{ec}$, this structure consists of the central part in the form of a circular superconducting part of the ceramics without a frozen field and a narrow ring around it, in which the entire frozen magnetic flux is concentrated. The external fields in the range of experimental values $10^2$–$5 \times 10^3$ A m$^{-1}$ correspond to frozen multiquantum magnetic fluxes containing from $10^4$ to $5 \times 10^5$ magnetic flux quanta. The outer diameter of the ring is approximately equal to the microsolenoid diameter, while the inner diameter depends on the properties of the superconductor and the magnitude of the frozen magnetic field. Superconducting currents with opposite directions, maintaining the frozen magnetic field, flow over the outer and inner circles of the ring. In the ceramic region located between them, the superconductivity of weak links between grains is suppressed by the frozen field. The central superconducting part of the structure is in the weakened magnetic field of the noted currents.

According to the model proposed, the initial linear portion of the dependence $H_{fm}(H_e)$ (Fig. 2, curve *2*) corresponds to freezing of the excitation field magnetic flux in the ring, which is proportional to the field magnitude. In this case, the magnetic flux freezing in the ring part of the ceramics is similar to the flux freezing in the annular gap between two coaxial superconductors. One of them, in the form of a solid circle, is at the center of the structure, while the other, with a hole, encircles it. In this case, the frozen magnetic field is also maintained by superconducting currents of opposite direction, which flow over the outer surface of the inner circle and the inner surface of the encircling superconductor.

After the external field reaches the value $H_{ec}$, the magnetic flux begins to penetrate the weakly coupled microcircuits in the central part of the structure and can be frozen there. The increase in the ring currents and the corresponding frozen magnetic field in the ring ceases at $H_e > H_{ec}$.

When a field close to $H_{ec}$ is frozen, it is necessary to take into account possible fluctuations of the sample temperature. In this case, at the same value of the excitation field exceeding $H_{ec}$, two alternative processes are possible. In one of them, the detected frozen magnetic field may increase proportionally to the increase in $H_e$ without penetration of the excitation field into the central part of the structure (a lower temperature). In the other process, the frozen magnetic field may somewhat decrease due to the penetration of the field $H_e$ into the central part of structure (a higher temperature). This phenomenon explains the existence of the transition region of ambiguous values of the frozen magnetic field in experimental curve *2* in Fig. 2. With a further increase in $H_e$, penetration of the field $H_e$ into the central part of the structure and its freezing in weakly coupled microcircuits in this ceramic region become dominant. The frozen fluxes are redistributed between the ring part of the structure (characterized by a strong magnetic coupling with the flux gate) and microcircuits in its central part (weaker magnetic coupling with the flux gate) in such a way that the output flux gate signal ceases to increase, and saturation of $H_f$ can be observed. This reconstruction of the current structure at $H_e = H_{ec}$ is a magnetic phase transition, which requires further study, in particular, using magnetic detectors with better spatial resolution in comparison with the flux gate.

The existence of the ring current structure is also confirmed by the investigations of the stability of the frozen magnetic field of this type to an additional local field. Let us consider a possible mechanism of its effect on the frozen magnetic field using the model of this current structure. The additional field, coinciding in direction with the frozen magnetic field, cannot affect the ring structure since this field, on the one hand, does not exceed $H_{ec}$ in the central part of the structure, and, on the other hand, is screened by the ring currents from penetration into the ring part of the structure. Vice versa, an additional field with the opposite direction decreases the magnetic flux in the ring, which may lead to the transition of some Josephson links in this part of the structure from the normal state. As a result, partial (limited by the critical current of these links) closing of superconducting ring currents through these weak links may occur, as well as their decrease. After switching off the additional field, a correspondingly lower magnetic flux will be frozen in the ring; i.e., the frozen magnetic field will decrease. With a further increase in $H_a$, the frozen field will decrease until the ring currents are reduced to zero and the frozen field is completely erased. An even stronger additional field will begin to penetrate the ceramics without a frozen field, as it occurs in the formation of a frozen magnetic field in the ZFC mode. In this case, the newly arising frozen field will coincide in direction with the additional field and be maintained by a radically different current structure in the form of a network of current microcircuits that is typical of the ZFC mode of the frozen field formation (Fig. 5a). The above-considered mechanism of the action of the field $H_a$ on the frozen magnetic field formed in the FC mode qualitatively explains the results obtained in the third-type experiments (Fig. 4, curve *2*).

## CONCLUSIONS

The dependences of the locally frozen magnetic field $H_{fm}$ in the $YBa_2Cu_3O_{7-x}$ ceramics with weak Josephson links between its grains on the excitation field $H_e$ have been obtained in a much larger range of $H_e$ values in comparison with our previous study [4]. The dependences obtained have a number of new features that have not been observed previously. They can be explained within the models of two current structures (for the FC and ZFC modes of frozen field formation) maintaining the frozen field, which were proposed in [4].

The portion with increasing frozen field in the dependence $H_{fm}(H_e)$ for the ZFC mode (Fig. 2, curve *1*) is determined by the range of critical currents of weak links in the ceramic region under study and can be used for contactless local diagnostics of the ceramic inhomogeneity on the micron and submicron levels. The smaller the difference between the external field $H_{es}$ corresponding to the onset of saturation of the frozen field and the critical field $H_{ec}$ corresponding to the onset of freezing, the higher the homogeneity of the ceramics with respect to the critical currents of its weak links. In addition, the magnitude of the frozen field in the saturation portion can be used for contactless estimation of the local maximum density of the critical current in the ceramics.

The dependence $H_{fm}(H_e)$ for the FC mode reveals a region of $H_e$ values corresponding to the frozen field values that are not reproduced from experiment to experiment. The model of the ring current structure for the FC mode suggests reconstruction of this structure in the noted region. This magnetic phase transition in superconducting ceramics has been observed for the first time and requires further investigation.

The changes in the magnitude of the local frozen field under the action of a demagnetizing dc field are also explained within the above-mentioned models of current structures and are another indirect proof of the existence of such structures. The dependences of the frozen field (ZFC mode) on the demagnetizing field suggest possibility of stable coexistence of microscopic magnetic fluxes with the opposite direction in the frozen-field region. These fluxes correspond to micrometer-sized Josephson current loops of the vortex–antivortex type. We can suggest that, in contrast to $YBa_2Cu_3O_{7-x}$ films [3], annihilation of such vortices and antivortices does not occur in the ceramics under study.